\documentclass[trackchanges,twocolumn]{aastex701}
\usepackage{float}

\begin{document}

\title{Investigating the Light Curves of 5 Long Period Variable Stars: A Study Cross-Harmonics using VStar, MESA, and GYRE}

\author[orcid=0009-0006-5677-3944,sname='Sherren']{Galina M. Sherren}
\affiliation{Institut Trottier de recherche sur les exoplanètes, Département de Physique, Université de Montréal, Montréal, Québec, Canada}
\email[show]{galina.sherren@umontreal.ca}

\author{Abigale Cross}
\affiliation{Department of Physics and Physical Oceanography, Memorial University of Newfoundland, 283 Prince Philip Drive, St. John’s, NL A1B
3X7, Canada}
\email[show]{abigale.cross@gmail.com}

\author[orcid=0000-0001-5744-4989,sname='Steenken']{Nicolaus Steenken}
\affiliation{Sternwarte Freimann, 80939 Munich, Germany}
\email[show]{nicolaus@steenken.info}

\author[orcid=0000-0002-7204-5502,sname='Ignace']{Richard Ignace}
\affiliation{Department of Physics \& Astronomy, East Tennessee State University, Johnson City, TN 37614, USA}
\email[show]{ignace@mail.etsu.edu}

\author[orcid=0000-0002-7322-7236, sname='Neilson']{Hilding R.~Neilson}
\affiliation{Department of Physics and Physical Oceanography, Memorial University of Newfoundland, 283 Prince Philip Drive, St. John’s, NL A1B
3X7, Canada}
\email[show]{hneilson@mun.ca}

\begin{abstract}
Long Period Variables (LPVs) are Asymptotic Giant Branch stars with periods $>$100 days. Around one third to half of LPVs display a Long Secondary Period (LSP) that is 5-10 times the length of the main periods. The exact cause of LSPs are unknown, with one potential explanation lying in cross-harmonics caused by the interaction of two simultaneously occurring periods. In this paper, we use the Fourier analysis software VStar to examine the visual range photometric light curves obtained through the American Association of Variable Star Observers of a sample of 5 stars. We find that one star, V CVn, displays a secondary period over 20 times the length of its main period, inconsistent with traditional LSPs. To test if cross-harmonics can result in an LSP or VCVn’s extra-long secondary period, we then construct evolutionary tracks and pulsation models of LPVs using Modules for Evolution in Stellar Astrophysics (MESA) and GYRE.  After examining the interactions of the fundamental pulsation mode and the first 4 overtones with each other, we do not find evidence that LSPs or V CVn’s extra-long secondary period could be caused by cross-harmonics. 

\end{abstract}


% Variable stars (1761), Intrinsic variable stars (859), Long period variable stars (935), Mira variable stars (1066), Multi-periodic variable stars (1079), Pulsating variable stars (1307), Multi-periodic pulsation (1078)
\keywords{Variable stars, Intrinsic variable stars, Long period variable stars, Mira variable stars, Multi-periodic variable stars, Pulsating variable stars, Multi-periodic pulsation}

% AAVSO; VStar; Photometry; AAVSO; V CVn; AK Peg; Z UMa; RX Boo; L$_2$ Pup; MESA; GYRE}

\section{Introduction}
\label{sec:intro}

Long Period Variable stars (LPVs) are a group of variable stars occupying the Asymptotic Giant Branch (AGB) \citep{catelan_pulsating_2015}. During this phase, LPVs undergo radial pulsations in the fundamental and first 3 overtones with main pulsation periods longer than 100 days. These red giant stars are formed from low-mass stars ( $\sim0.6$ to $ \sim 8\ M_{\odot}$), thus representing the future of our sun \citep{neilson2023multiyear,catelan_pulsating_2015}.

LPVs are made up of two categories: Mira and Semi-regular Variable (SRV) stars, where Mira variables are more evolved than SRVs \citep{catelan_pulsating_2015}. While these two categories have distinct features, there is no clear boundary as to where SRVs stop and Miras begin. The main differences lie in the regularity of their pulsation, and the amplitude of the brightness variation \citep{catelan_pulsating_2015}. \cite{Wood_2000} determined that Miras pulsate in the radial fundamental mode, and SRVs pulsate in the radial fundamental and first three overtone modes. LPVs together form one of the largest unknowns in current-day stellar astrophysics. As they are evolved low-mass variable stars, their mostly convective envelopes in the AGB could cause interactions between pulsation and convection that are not  yet well understood \citep{catelan_pulsating_2015}. 

Approximately one-third of LPVs display a long secondary period (LSP): a secondary period that is about 5-10 times longer than the main pulsation period \citep{Wood_2000,percy2016studies}. To date there is no definite explanation as to the cause of LSPs, although distinct period-luminosity relations for LPVs with LSPs have been identified \citep{Wood_2000,percy2016studies}. The most recently proposed explanations involve a dusty orbiting cloud \citep{soszynski2021binarity} or a binary companion as in the case of Betelgeuse \citep{goldberg2024buddy}.

\cite{Neilson_2014} examined the polarimetric variability of the SRV star V Canum Venaticorum (V CVn), and determined that the star emits between 1-8 $\%$ linear polarized light anti-correlated with its brightness. They proposed a cause for this behaviour in the form of a variable bow shock as the star travels through the interstellar medium \citep{Neilson_2014, neilson2023multiyear}. Alongside V CVn, a sample of stars with similar qualities and polarization behaviour were identified: L$_2$ Pup, RX Boo, AK Peg, Z Uma, and UZ Ari . Despite multiple theories about the cause of this phenomenon, an analysis of the long-term light curves of these stars has not yet been performed.

In this work, we analyze over a century of visual photometric data from the American Association of Variable Star Observers (AAVSO) using Fourier and Wavelet analysis techniques to extract information on the stars in that sample as well as V CVn. UZ Ari is omitted as the star does not have sufficient observations. We use the program VStar by the AAVSO to perform our analysis \citep{benn2012algorithms+}. As a follow-up to our results, we then use ``Modules for Evolution in Stellar Astrophysics" (MESA) to construct a model for low-mass LPVs to then test a theory of cross-harmonics using GYRE.

\section{Methods} \label{sec:style}

We used visual photometric data obtained through the AAVSO for our analysis \citep{aavso_data}. This data is imported in the form of a time series into the analysis software VStar \citep{benn2012algorithms+}. VStar is an open-source time series analysis and visualization tool developed by David Benn primarily intended for variable star light curves \citep{benn2012algorithms+}. It is compatible with data from the American Association of Variable Star Observers (AAVSO) and allows for significant periods of pulsation to be extracted and verified though a variety of methods \citep{benn2012algorithms+}. 

\begin{figure*}[t]
    \centering
    \includegraphics[width=\textwidth]{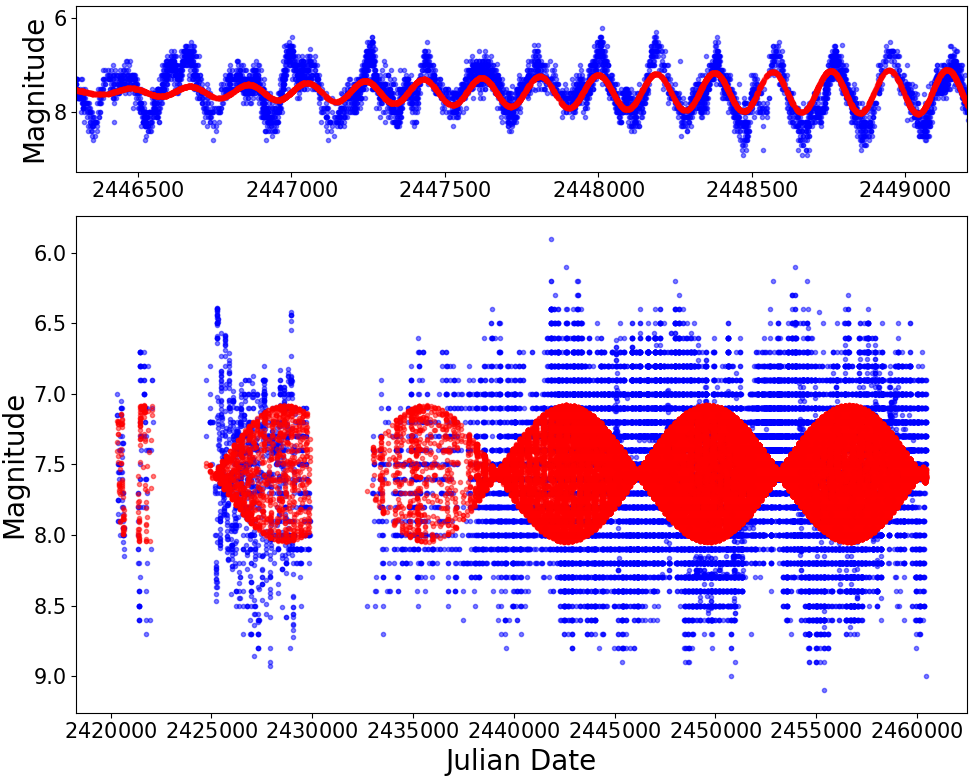}
    \caption{AAVSO visual photometric data of V CVn (blue) fit with a model of $194$ and $186$ day periods (red) obtained through VStar. The top plot displays a $\approx~10$~year region of pulsation taken from the bottom plot. The two large gaps correspond to 5-7 years each of no observations, likely due to World War I and World War II. The blue dots correspond to AAVSO data, while the red dots are a model fit of two periods of 194 and 186 days, which is the most recent agreement in the literature \citep{kiss1999multiperiodicity,kiss2000multiperiodicity,samus2017general}.}
    \label{fig:VCVNObs+Model}
\end{figure*}

The data for V CVn and the sample of stars is irregularly sampled, spanning about 100 years, and is obtained by different observers at different locations. The AAVSO carries out normalization and data verification methods, and as such only data with a ``high quality" designation was used. The visual photometric AAVSO data for VCVn is shown in Fig. \ref{fig:VCVNObs+Model}. Some AAVSO data in the lower plot appears discretized at intervals, likely due to normalization methods used by AAVSO and the observers. The larger discretization gaps are 0.1 mag, while other more recent ones occur within 0.01 mag of each other. The top plot of Fig. \ref{fig:VCVNObs+Model} displays a region of observations spanning around 10 years. Some pulsations display double peaks, which could indicate double periodicity.

We first employ VStar's Date Compensated Discrete Fourier Transform (DC DFT) developed by Ferraz-Mello as our primary period extraction technique \citep{ferraz1981estimation}. The algorithm accounts for unequally spaced data sampling, and yields a period with a corresponding power and semi-amplitude for each trial period \citep{benn2012algorithms+}. The power indicates the statistical significance of a given trial period's fit to the data \citep{benn2012algorithms+}. As a result, we obtain suggestions for the most probable periodic behaviour in the star. All uncertainties reported in the frequency analysis are the full width half maximum (FWHM) values of said frequency powers. The photometric data does not include an error calculation from the AAVSO or the observers, and thus cannot be commented on. 

%%%%%%% AoV %%%%%%%%

Once a power spectra is obtained through the DC DFT, significant periods can be refined using the CLEANest algorithm \citep{foster1995cleanest}. The irregular spacing of data can cause minor frequency alterations and major amplitude changes in the power spectra, while also introducing false peaks \citep{foster1995cleanest}. We apply CLEANest to the significant periods of the stars to determine the true frequencies.

%%%%%% WWZ %%%%%%%%

We also apply the Weighted Wavelet Z-Transform routine to the light curves, which is carried out with the VStar WWZ function \citep{foster1996wavelets,benn2012algorithms+}. The WWZ is a weighted least-squares fit that allows the visualization of how the periodic pulsation might change over time \citep{foster2010analyzing}. The result is a 3D contour plot of where given periods became significant during a star's lifetime.

%%%%%%%%% AoV %%%%%%%%%%

A second method of period analysis was used through VStar, which uses the One-Way Analysis of Variance (AoV) algorithm to compute an F-statistic for corresponding periods. We combine the analysis of the raw data using the DCDFT with residual analysis after an initial model is fit. This allows for the determination of aliased and cross-harmonic frequencies in the DCDFT spectrum \citep{foster2010analyzing}. While aliasing is an artefact of Fourier analysis, cross-harmonics are a real physical phenomenon, and have been observed in stars such as TU Cas \citep{foster2010analyzing}.

For V CVn, L$_2$ Pup, RX Boo, AK Peg, and Z UMa, we first apply a DCDFT to the unaltered AAVSO visual photometric data. We then apply a combination of the DCDFT, AoV, and WWZ analysis methods to binned data and the residual data after a model is fit. The individual conditions of the stars required different analysis approaches, and are discussed in the results. We then move on to test if multiple-mode LPV pulsation that leads to cross-harmonic interaction has the potential to cause an LSP.

%%%% Cross-Harmonics %%%%%%%%

Cross-harmonics, as described by Foster (2010), are the sum of the sinusoids of the main frequencies \citep{foster2010analyzing}. This can be described using the equation:

\begin{equation}
    |\nu_{12}| = j\nu_1 + k\nu_2,
    \label{eq:crossHarm}
\end{equation}
 where j and k are any positive or negative integer and $\nu_1$ and $\nu_2$ are simultaneously occurring periods \citep{foster2010analyzing}. 
 \

%%%%%%%% MESA %%%%%%%%%%%
In order to apply the cross-harmonic formula, we first construct stellar evolution models of 1.0, 1.1, 1.2, 1.3 and 1.5 $M_{\odot}$, for which we then calculate the pulsation frequencies in the AGB regimes. This range of mass was selected for their abundances in the initial mass function, and time limit of the study. We use ``Modules for Experiments in Stellar Astrophysics" (MESA) version r23.05.1, which is a FORTRAN 95 based open-source code developed to make stellar models, and can model a variety of stellar phenomena \citep{paxton_modules_2011,paxton_modules_2013,paxton_modules_2015,paxton_modules_2018,paxton_modules_2019,jermyn_modules_2023, paxton_2023_7983526}. We utilize MESA to calculate the evolution tracks of the stars with the desired initial mass. We then use GYRE to calculate their pulsation frequencies in the AGB phase \citep{townsend_gyre_2013}. GYRE is MESA compatible and solves the fundamental stellar pulsation equations using a Magnus Multiple Shooting scheme for solving linearized pulsation equations, and is written in FORTRAN 2008 \citep{townsend_gyre_2013}. Combining these methods, we calculate the model's pulsation frequencies for the fundamental mode and first 4 overtones. 

We then apply the first-order cross-harmonic equation between all the frequencies that occur simultaneously to determine if some LSPs or extra-long secondary periods could be caused by cross-harmonics. The first-order cross-harmonic is taken to be the $j,k = 1$ case of Equation \ref{eq:crossHarm}. This is also consistent with Kiss's (2000) argument of ``cross-production" in V CVn \citep{kiss2000multiperiodicity}.

The evolution tracks are based off MESA's test suite run that evolves a 1-$M_{\odot}$ star from main sequence to white dwarf. The tracks terminated at the end of the thermally pulsing AGB (TP-AGB) phase, and the timestep resolution during the AGB phase was increased. Snapshots during the AGB phase were then used to calculate the pulsation periods using GYRE \citep{townsend_gyre_2013}. An adiabatic assumption was made during the period calculation. Although a non-adiabatic calculation was attempted, the result was indicative of a possible bug or inability to converge on the correct solution. An example of a non-adiabatic solution is given in appendix \ref{app: NASolution}. As the main purpose of a non-adiabatic calculation is to determine dampening rates of pulsation modes, this is expected to not have a significant impact on our results.

\section{Results} \label{sec:floats}

\subsection{V CVn}

\cite{kiss2000multiperiodicity} first reported the possibility of cross-harmonic periods in V CVn by analyzing the light curve of the star using Fourier analysis. We repeat this process with the added photometry of the last 24 years. 

\begin{figure}[h]
    \centering
    \includegraphics[width = \columnwidth]{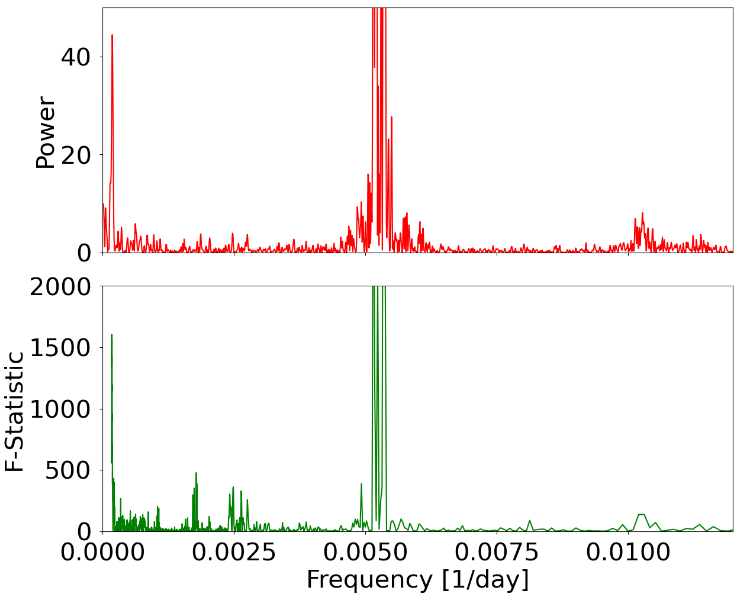}
    \caption{DCDFT of 10-day binned data of V CVn (top), and
    AoV with 1-day bins (bottom).}
    \label{fig:VCVnAnalysis}
\end{figure}

\begin{figure}[h]
    \centering
    \includegraphics[width = \columnwidth]{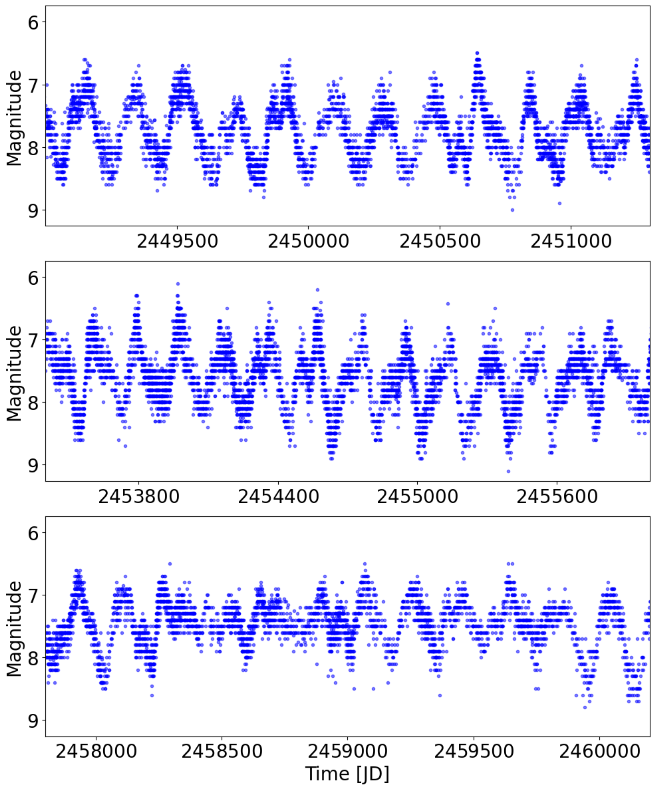}
        \caption{The light curve of V CVn displayed in three regions. The amplitude and period variability is apparent, as well as double-peaks during multiple maxima.}
    \label{fig:VCVnZoomed}
\end{figure}

As shown in Fig. \ref{fig:VCVnAnalysis}, we find a main period of 192 $\pm$ 0.4 days with an amplitude of 0.44 $\pm$ 0.02 mag corresponding to a frequency of $\sim 0.005 \ d ^{-1}$ followed by a 186 $\pm$ 0.4 $d$ period with a 0.43 $\pm$ 0.01 mag amplitude in the same region. In V CVn's analysis and the following stars, the amplitude analyses have a larger margin of error than the period length, likely caused by variability in the periods and data. This is also demonstrated by the noise surrounding the peaks at $\sim0.005 \ d ^{-1}$. As discussed by \cite{kiss2000multiperiodicity} and \cite{buchler2004evidence}, this indicates multiple periods and is listed as such in \cite{samus2017general}. The multiperiodic nature is also supported by the light curve: certain points in the light curve display double peaks, as shown in Fig. \ref{fig:VCVnZoomed}. Also indicated in the DCDFT spectrum are multiple secondary peaks flanking the main period. The frequencies located at $\sim$0.0025 and $\sim$0.008 $day^{-1}$ are aliases of the yearly variability of 0.002738 $cyc/d$ \citep{foster2010analyzing}. The 1-year alias appears as flanking peaks and is caused by the data density that is split every year by the sun blocking the view of the star, and can be described as:

\begin{equation}
    \nu \pm 0.002738
\end{equation}
where $\nu$ is the main period(s). The two flanks match within $\pm$0.0001 of the one-year aliasing. The next secondary peak is at ~0.013 $d^{-1}$. The most likely explanation is ``cross-harmonics" as given in Eq. \ref{eq:crossHarm}, as the frequency is equivalent to the sum of the two main peaks. These three peaks diminish when 10 day bins are applied as in Fig. \ref{fig:VCVnAnalysis}.

We also find a longer period at 5460 days with an amplitude of 0.22 $\pm$ 0.02 mag using the DC DFT Method followed by refinement using the CLEANest algorithm. This period is especially visible in the middle plot of Fig. \ref{fig:VCVnAnalysis} at a value of 0.000184 $\pm$ 0.000026 $d^{-1}$, or 5441 $\pm$ 680 $d$. A previous period analysis on V CVn by \cite{kiss2000multiperiodicity} attributes this 5460-day period to ``cross-production" caused by the subtraction of double pulsation, as argued for the peak at ~0.013 $d^{-1}$ \citep{kiss2000multiperiodicity}. We assume that ``cross-production" as in \cite{kiss2000multiperiodicity} is referencing the same phenomenon that we describe in this paper. This 5460-day period is hereafter referred to as the ``Extra Long Secondary Period", or ELSP, as it is much longer than most measured LSPs \citep{Wood_2000,percy2016studies}. One would expect both cross-harmonic peaks to be of equal or close powers, yet this is not the case for the ELSP. The ELSP is around 3 times more powerful than the other cross-harmonic peak in the raw data DCDFT, and around 15 times stronger in the binned DCDFT. We also observe a strong possibility of rejecting the null hypothesis at the ELSP from the AoV spectrum, which is around 100 times stronger than the peak at ~0.013 $d^{-1}$ as shown in Fig. \ref{fig:VCVnAnalysis}. Finally, we apply a WWZ analysis to the determined periods, but yield no valuable conclusions as the desired resolution could not be achieved with VStar. This is the case for all stars analyzed, and as such WWZ analysis is left out of our results.
\subsection{RX Boo}

\begin{figure}[h]
    \includegraphics[width = \columnwidth]{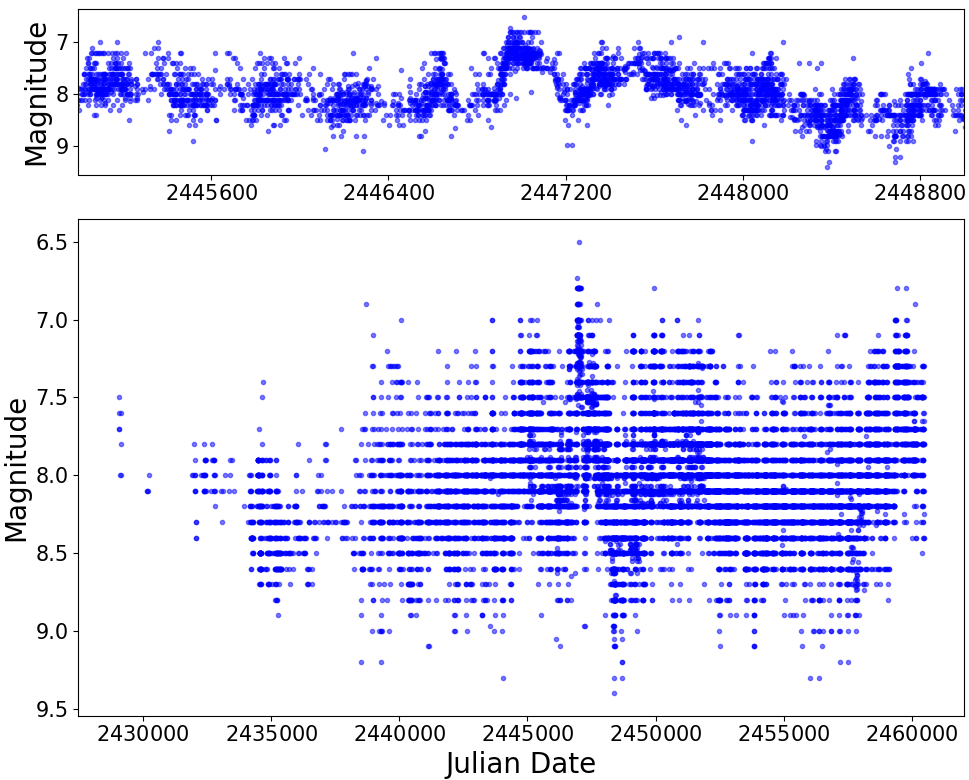}
    \caption{AAVSO visual photometric data of RX Boo (bottom) with a select timespan of variability (top).}
    \label{fig:RXBooPhotometric}
\end{figure}

RX Boo is a SRb star with a well-studied main period of 161$\pm$1 and an LSP of 2205.1 d \citep{percy2016studies, samus2017general, neilson2023multiyear, cadmus2024long}. The AAVSO visual range (see Fig. \ref{fig:RXBooPhotometric}) photometric data spans from 1938 to 2024, with the first $\sim$30 years being sparse.

\begin{figure}[h]
    \centering
    \includegraphics[width = \columnwidth]{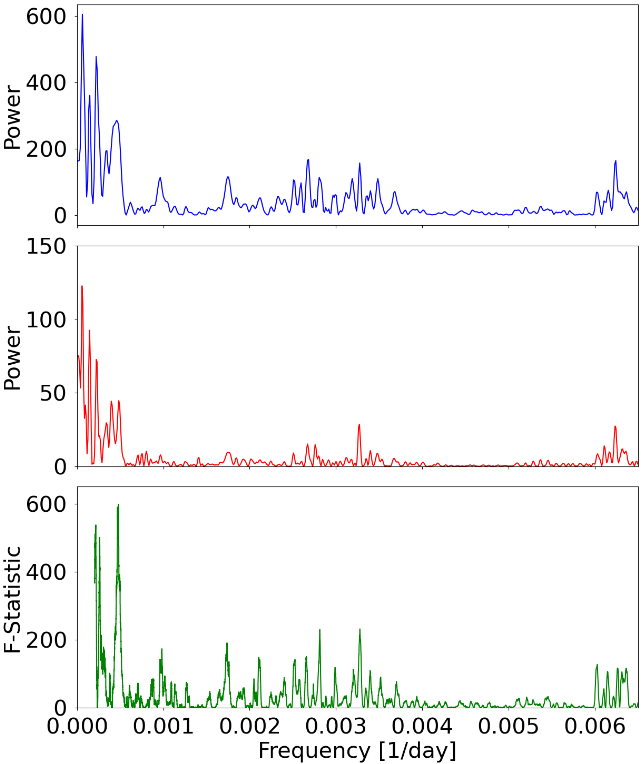}
    \caption{RXBoo DCDFT of raw visual photometric data (top), DCDFT of 10-day binned data (middle), and AoV with 1-day bins (bottom).}
    \label{fig:RXBooAnalysis}
\end{figure}

As a result of the DCDFT, we find the main pulsation period of 160 $\pm$ 0.3 d (0.006242 $\pm$ 0.000012) with an amplitude of 0.124 $\pm$ 0.004 mag as shown in Fig. \ref{fig:RXBooAnalysis} and verified using CLEANest. We also find an LSP of 2186 $\pm$ 261 d with a 0.170 $\pm$ 0.003 mag amplitude or 2205 d when refined using the CLEANEST algorithm, agreeing with GCVS data \citep{samus2017general}.

The LSP region is quite noisy, with the top peak in the DCDFT and AoV spectrums corresponding to 16181 $\pm$ 3800, 4494 $\pm$ 3800, and 6742 $\pm$ 492 d. The 4494 and 6742 day periods are likely a harmonic of the LSP. The 16181-day period is of questionable legitimacy as the time span of the data is $\sim$ 31000 d, i.e. only around 2 periods. The peaks at ~0.0028 correspond to the 1-year aliasing peaks, and sharply displayed in the middle plot of Fig. \ref{fig:RXBooAnalysis}. As such, we do not find evidence of an ELSP in RX Boo.

\subsection{$L_2$ Pup}

 L$_2$ Pup is an oxygen-rich SRb star with a complex circumstellar medium \citep{kervella_alma_2016,lykou_dissecting_2015,bedding2002light}. Polarization measurements indicate a polarization variability between 0 to 8 \%, with a constant PA of 165-180 degrees \citep{neilson2023multiyear,magalhaes1986polarimetric}. Earlier studies on $L_2$ Pup's light curve indicate a period of approximately 140 days with moderate variability \citep{bedding2002light}. 
The star is surrounded by a near edge-on dusty torus with inner radius 4.2 AU, the likely cause of a great dimming event starting in 1995 \citep{kervella_alma_2016,bedding2002light}. 
L$_2$ Pup is thought to be orbited by a companion dubbed L$_2$ Pup B at 2.43 $\pm$ 0.16 AU, likely a compact body \citep{kervella_alma_2016}. \cite{kervella_alma_2016} calculated the orbital period of the companion to be 4.69 $\pm$ 0.45 years. The AAVSO data for L$_2$ Pup in the visual range spans 1891 to 2024.
 
\begin{figure}[h]
    \centering
    \includegraphics[width = \columnwidth]{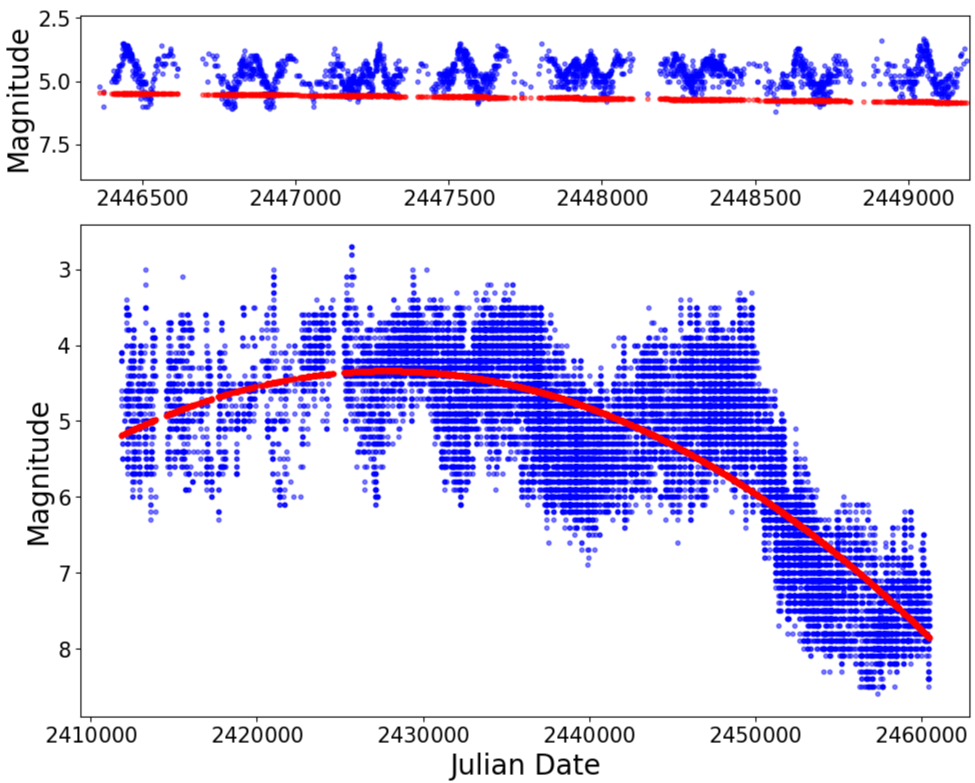}
    \caption{AAVSO visual photometric data of L$_2$ Pup (blue) (bottom) and a select region of pulsation (top) fitted with a model accounting for the dimming event (red).}
    \label{fig:L2PupDimFit}
\end{figure}

\begin{figure}[h]
    \centering
    \includegraphics[width = \columnwidth]{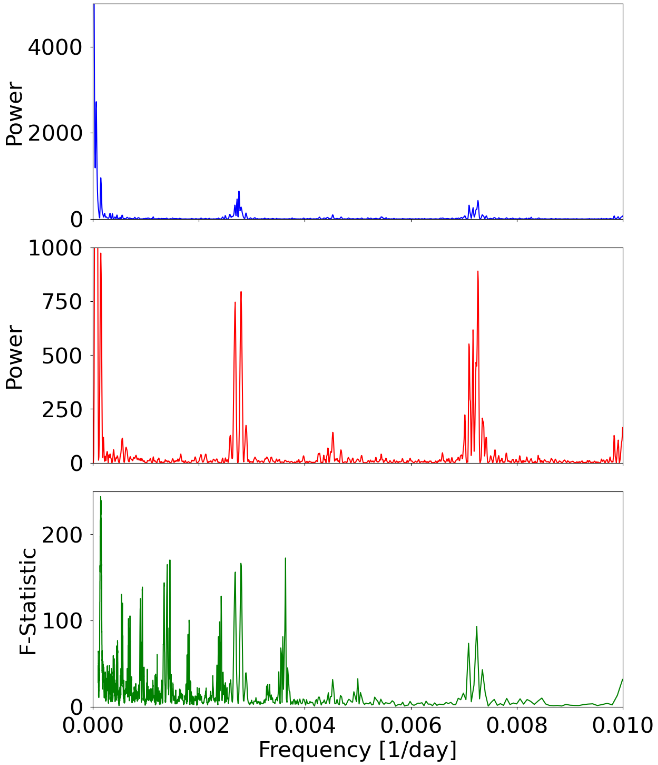}
    \caption{L$_2$ Pup DCDFT of raw visual photometric data (top), DCDFT of residual data from fitting the dimming curve (middle), and AoV with 10-day bins (bottom).}
    \label{fig:L2PupAnalysis}
\end{figure}

Due to the dimming event of $L_2$ Pup occuring beginning around May 1990, VStar picked out the strongest peak to fit the dimming curve. This was at a period of 142272 d, which is larger than our data range of 48910 d. We fit the data with this period model, then subtract from the light curve to find the residuals. This fit is illustrated in Fig. \ref{fig:L2PupDimFit}. The DCDFT on the raw and binned data as well as the AoV was then applied to this subtracted data. The top plot in Fig. \ref{fig:L2PupAnalysis} which displays the raw DCDFT without subtracting the dimming event displays how powerful the peak for the dimming is.

As a result of running the DC DFT on the residual light curve and applying the CLEANest algorithm to the significant periods, we find a main period of 139.3 $\pm$ d with an amplitude of 0.40 $\pm$ 0.06 mag and a much longer period at 6780 $\pm$ 600 d with an amplitude of 0.57 $\pm$ 0.01 mag. The 140-d region of the DCDFT is noisy, indicating variability in the main period or poor data sampling. While the main period agrees with past literature on $L_2$ Pup, we were unable to find note of the longer, smaller amplitude period at 6780 $\pm$ 600. As it shows similar properties to the ELSP of V CVn, such as the length ratio of the secondary period to the main period and the amplitude, we will also refer to this period as an ELSP. We can further confirm the legitimacy of the ELSP using the AoV, where the strongest peak corresponds to the 6780 $\pm$ 600 d period. The two flanks at $\sim$0.0044 and $\sim$0.009 $d^{-1}$ are the 1-year aliasing peaks. We also find a period of 353 $\pm$1 with a 0.4 $\pm$ 0.01 mag amplitude, however the noise in this region is high, and the actual uncertainty range may be higher than the FWHM. The source of this period is unclear as it is not in the range of a harmonic or a cross-harmonic as there is no confirmed multiperiodic behaviour.

\subsection{AK Peg}
\begin{figure}[t]
    \centering
    \includegraphics[width=\columnwidth]{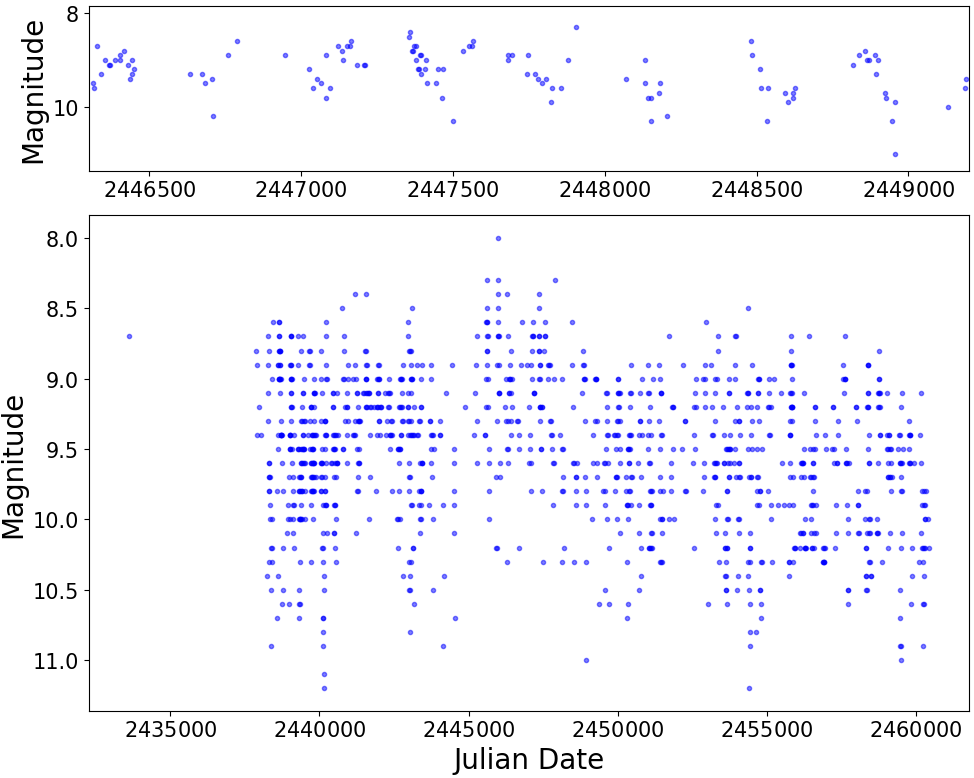}
    \caption{All AAVSO visual photometric data of AK Peg (bottom), with a selected timespan of variability (top).}
    \label{fig:AKPegPhotometric}
\end{figure}

AK Peg visual photometric data spans 1950 to 2024, and is sparsely collected as demonstrated in Fig. \ref{fig:AKPegPhotometric}. The star has a main period of 194 d, with no noted LSP \citep{samus2017general}. 

\begin{figure}[h]
    \centering
    \includegraphics[width = \columnwidth]{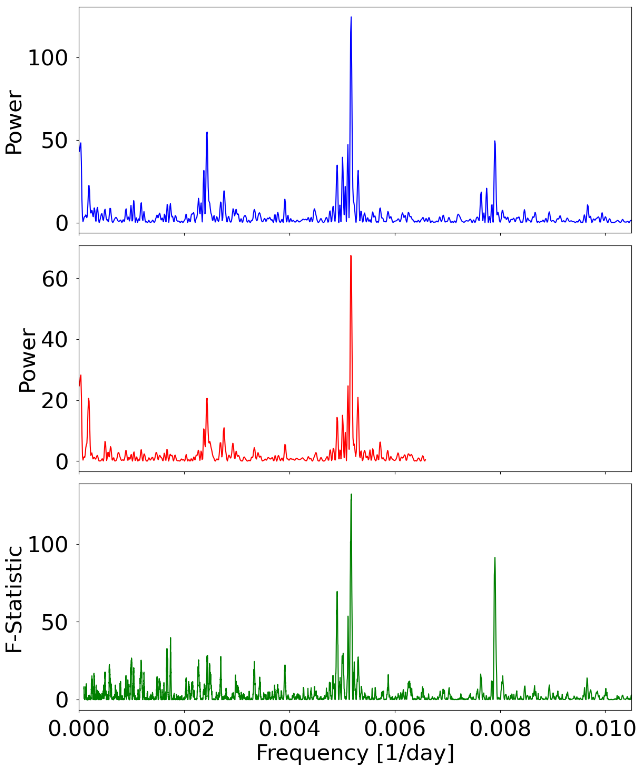}
    \caption{AK Peg DCDFT of raw visual photometric data (top), DCDFT of 10-day binned data (middle), and AoV with 1-day bins (bottom).}
    \label{fig:AKPegAnalysis}
\end{figure}

 From the analysis, we determine a 193.2 $\pm$ 0.6 d period with the DCDFT, and a 193.3 d period using CLEANest. The results of the DCDFT for the raw data, 10-day bins, and the 1-day binned AoV is given in Fig. \ref{fig:AKPegAnalysis}. The peaks at 0.0025 and 0.008 $d^{-1}$ correspond to the 1-year alias. The peak at near-zero corresponds to a period of $\sim$24000 d. Considering the data ranges 26000 d, this period cannot be verified and is likely an artefact of the data analysis. AK Peg has no noted LSP in literature, which agrees with our analysis.

\subsection{Z Uma}

Z UMa is an SRb star with visual photometric AAVSO data spanning 1934 to 2024. It is a well-observed star with a light curve indicating a possible double-periodicity at 195 and 203 d, as well as a longer period at 4200 d \citep{suchko1980periodicities}. \cite{suchko1980periodicities} argue that a 4200 d period is caused by a beating of these two closely-spaced periods \citep{suchko1980periodicities}.

\begin{figure}[h]
    \centering
    \includegraphics[width = \columnwidth]{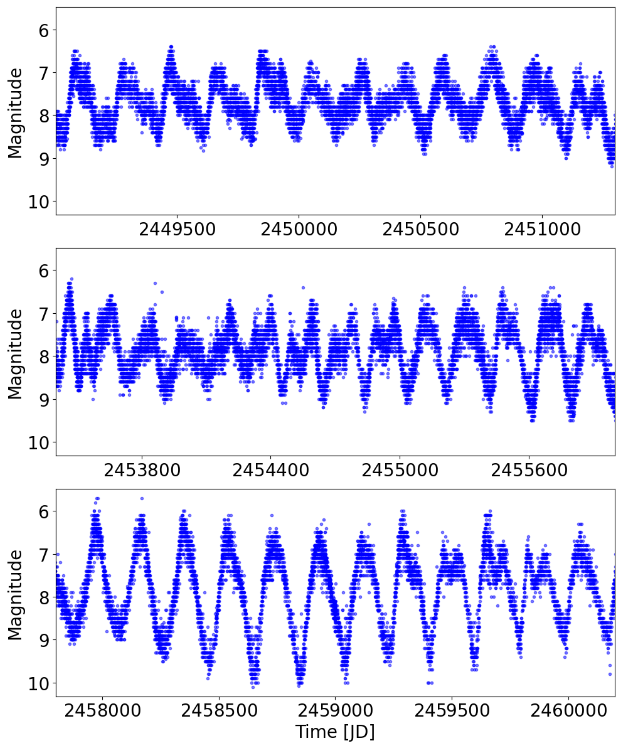}
    \caption{Z UMa AAVSO visual photometric light curve in three select regions of earliest to latest in descending order.}
    \label{fig:ZUMaZoomedLC}
\end{figure}

 In our analysis, this double-periodic behaviour is supported by the light curve of Z UMa. In the zoomed-in light curve in Fig. \ref{fig:ZUMaZoomedLC}, double-peaks appear at certain points, which are indicative of double periods.

\begin{figure}[h]
    \centering
    \includegraphics[width = \columnwidth]{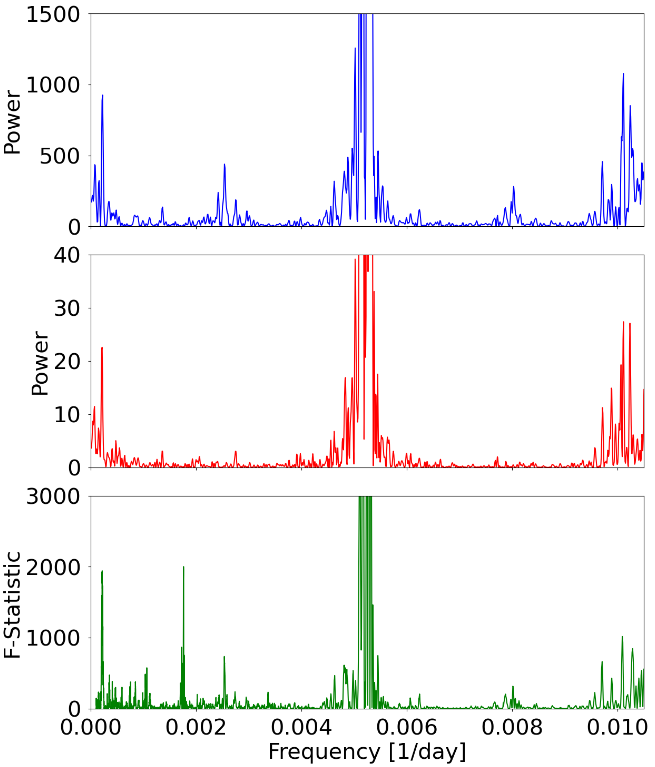}
    \caption{Z Uma DCDFT of raw visual photometric
    data (top), DCDFT of 10-day binned data (middle), and
    AoV with 1-day bins (bottom).}
    \label{ZUmaAnalysis}
\end{figure}

We determine a 189 $\pm$ 1 d period with a amplitude of 0.30 $\pm$ 0.04 mag, alongside a 193 $\pm$ 0.6 d period with an amplitude of 0.6 $\pm$ 0.02 mag. These values are not in agreement with \cite{suchko1980periodicities} or \cite{samus2017general}, which likely due to data quality or the time variation of these periods. We also find the 1-year alias peaks at $\sim$0.002 and $\sim$0.008 $d^{-1}$, as well as two cross-harmonic peaks at $\sim$0.0002 and $\sim$0.01 corresponding to 4668 $\pm$ 223 and 99 $\pm$ 1 d. Different from V CVn, the cross-harmonic peaks caused by the double periodicity are within 17$\%$ power of each other. The determined periods of all 5 stars are summarized in Table \ref{table1}.

%% The values (usually only l,r and c) in the last part of
%% \begin{deluxetable}{} command tell LaTeX how many columns
%% there are and how to align them.
\begin{deluxetable*}{ccccc}

%% Keep a portrait orientation

%% Over-ride the default font size
%% Use Default (12pt)

%% Use \tablewidth{?pt} to over-ride the default table width.
%% If you are unhappy with the default look at the end of the
%% *.log file to see what the default was set at before adjusting
%% this value.

%% This is the title of the table.
\tablecaption{Summary of periods determined in this paper}

%% This command over-rides LaTeX's natural table count
%% and replaces it with this number.  LaTeX will increment 
%% all other tables after this table based on this number
\tablenum{1}
\label{table1}
%% The \tablehead gives provides the column headers.  It
%% is currently set up so that the column labels are on the
%% top line and the units surrounded by ()s are in the 
%% bottom line.  You may add more header information by writing
%% another line between these lines. For each column that requries
%% extra information be sure to include a \colhead{text} command
%% and remember to end any extra lines with \\ and include the 
%% correct number of &s.
\tablehead{\colhead{Star} & \colhead{Period 1} & \colhead{Period 2} & \colhead{Long Secondary Period} & \colhead{Extra-Long Secondary Period} \\ 
\colhead{} & \colhead{(Days)} & \colhead{(Days)} & \colhead{(Days)} & \colhead{(Days)} } 

%% All data must appear between the \startdata and \enddata commands
\startdata
V CVn & 192 & 186 & N/A & 5600\\
RX Boo & 160 & N/A & 2205 & N/A\\
$L_2$ Pup & 139 & N/A & N/A & 6780\\
AK Peg & 193 & N/A & N/A & N/A\\
Z UMa & 189 & 193 & N/A & N/A\\
\enddata

%% Include any \tablenotetext{key}{text}, \tablerefs{ref list},
%% or \tablecomments{text} between the \enddata and 
%% \end{deluxetable} commands

%% No \tablecomments indicated

%% No \tablerefs indicated

\end{deluxetable*}

\section{1D Stellar Models}

\begin{figure}[h]
    \centering
    \includegraphics[width = \columnwidth]{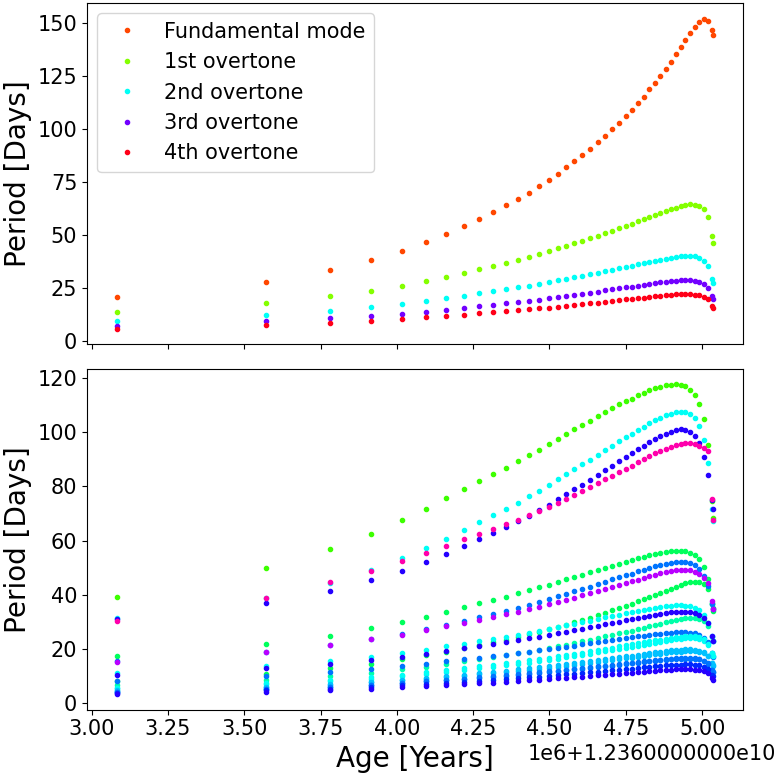}
    \caption{1 $M_{\odot}$ pulsation periods in the fundamental, and 1st, 2nd, 3rd, and 4th overtones (top), and the cross-harmonic calculation for simultaneously occuring periods (bottom).}
    \label{fig:1MPlot}
\end{figure}

\begin{figure}[h]
    \centering
    \includegraphics[width = \columnwidth]{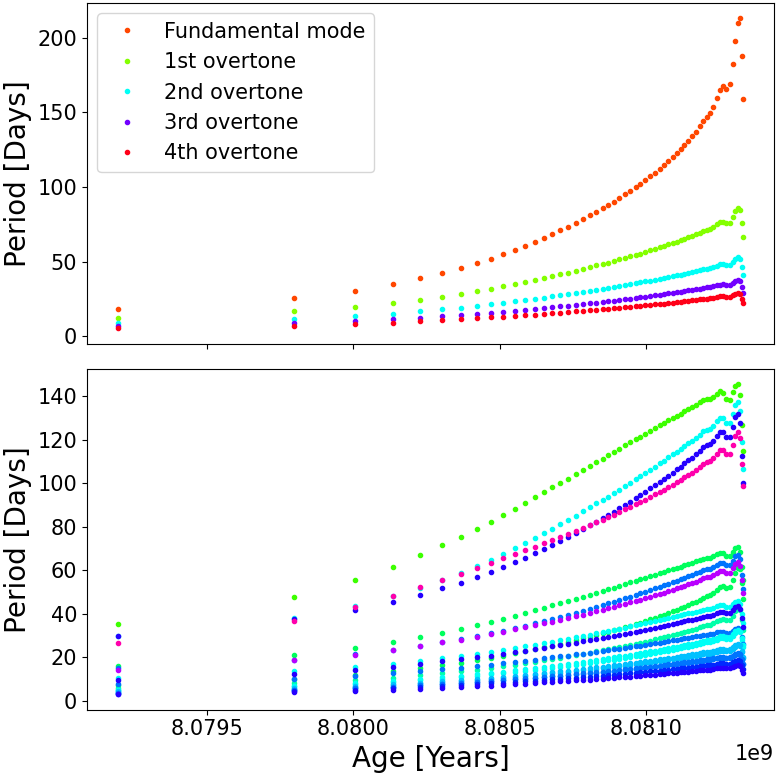}
    \caption{1.1 $M_{\odot}$ pulsation periods in the fundamental, and 1st, 2nd, 3rd, and 4th overtones (top), and the cross-harmonic calculation for simultaneously occuring periods (bottom).}
    \label{fig:1.1MPlot}
\end{figure}

\begin{figure}[h]
    \centering
    \includegraphics[width = \columnwidth]{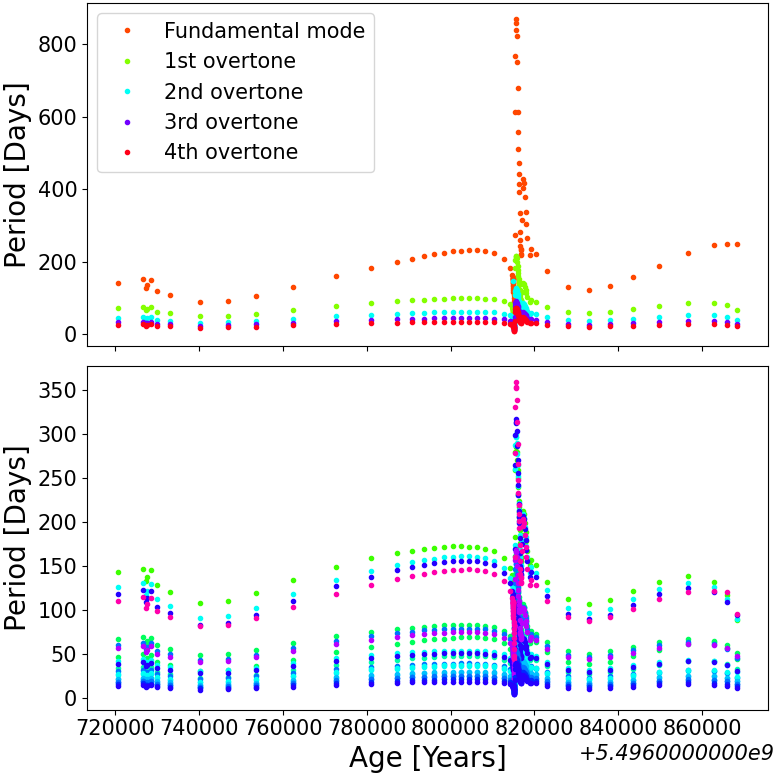}
    \caption{1.2 $M_{\odot}$ pulsation periods in the fundamental, and 1st, 2nd, 3rd, and 4th overtones (top), and the cross-harmonic calculation for simultaneously occuring periods (bottom).}
    \label{fig:1.2MPlot}
\end{figure}

\begin{figure}[h]
    \centering
    \includegraphics[width = \columnwidth]{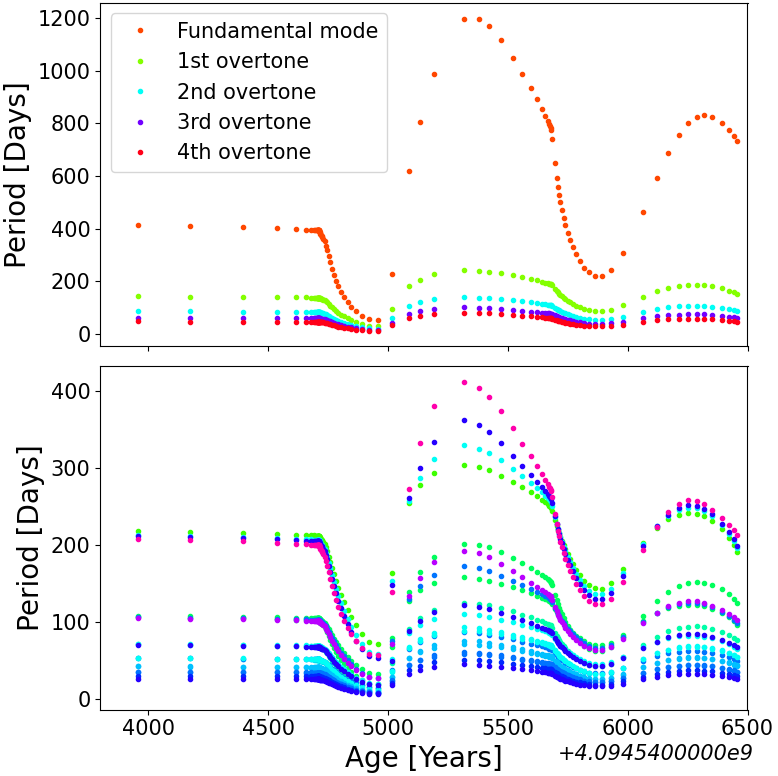}
    \caption{1.3 $M_{\odot}$ pulsation periods in the fundamental, and 1st, 2nd, 3rd, and 4th overtones (top), and the cross-harmonic calculation for simultaneously occuring periods (bottom).}
    \label{fig:1.3MPlot}
\end{figure}

\begin{figure}[h]
    \centering
    \includegraphics[width=\columnwidth]{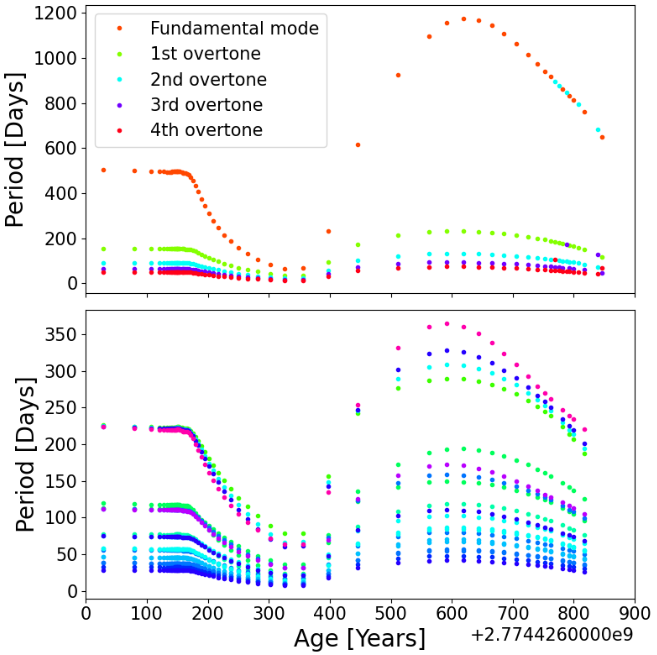}
    \caption{1.5 $M_{\odot}$ pulsation periods in the fundamental, and 1st, 2nd, 3rd, and 4th overtones (top), and the cross-harmonic calculation for simultaneously occuring periods (bottom).}
    \label{fig:1.5MPlot}
\end{figure}

Using the constructed stellar evolution models paired with the pulsation period extraction, we test if two closely-spaced periods as in V CVn and Z UMa carry the potential to cause the period beating phenomenon. For models of 1.0, 1.1, 1.2, 1.3 and 1.5 $M_{\odot}$, we display the periods during the AGB phase in Fig. \ref{fig:1MPlot} through Fig. \ref{fig:1.5MPlot}. For the 1.2 and 1.3 $M_{\odot}$ models, the central rapid period changes are identified as thermal pulses, and follows the shape of a thermal pulse from \cite{joyce2024stellar}.  

Outside of the thermal pulse regions, no model displays two periods within 20 days of each other as in Z UMa or V CVn for the fundamental mode and first, second, and third overtones. In no model do we observe the cross-harmonic calculation to result in a period 5-10 times longer than the main pulsation period for any pulsation mode. The thermally pulsing regions where rapid period changes occur do display closely-spaced periods, but the resulting first-order cross-harmonics do not resemble an ELSP such as in V CVn or Z UMa. As such, we conclude that the stellar evolution and pulsation models we use do not predict an ELSP caused by cross-harmonics.

\section{Discussion}

V CVn's ELSP at 5441 $\pm$ 680 days has been referred to as a cross-harmonic peak by \cite{kiss2000multiperiodicity} through the subtraction of the two main periods of the star. Our findings are in contention with the ELSP being caused from cross correlation, due to the asymmetric nature of the cross production terms, as well as the DC DFT spectrum using 10-day binned averages. The 1-year aliasing peaks, as well as the cross-harmonic peaks proposed by \cite{kiss2000multiperiodicity} disappear except for the proposed ELSP in the binned DC DFT. This case is supported by the light curve of Z UMa which displays two near-equal cross-harmonic peaks that are close in power. There is no clear reason for V CVn's asymmetric cross-harmonic peaks, which could indicate a different extrinsic or intrinsic process that is causing a strengthening of this period.

One possible explanation of V CVn's ELSP is that of long secondary periods (LSPs). The ELSP we determine is over 20 times longer than the main periods of the star, thus making it either unlikely to be an LSP, or one of the longest main period to LSP ratios discovered thus far. 

We would also like to acknowledge shortcomings of our pulsation models using MESA and GYRE that can be addressed in future work. Given that masses higher than 1.5  $M_{\odot}$ have not been tested, this study is not a conclusive result. We also acknowledge that the interplay between convection and pulsation in LPVs is not a well-understood process, and different models may result in different pulsation period ratios \citep{catelan_pulsating_2015}. This portion of our paper serves to demonstrate that this type of modeling and calculation is possible using MESA and GYRE, and can be successfully compare to the light curves of LPVs. 

\begin{figure*}[!htbp]
    \centering
    \includegraphics[width = 0.95\textwidth]{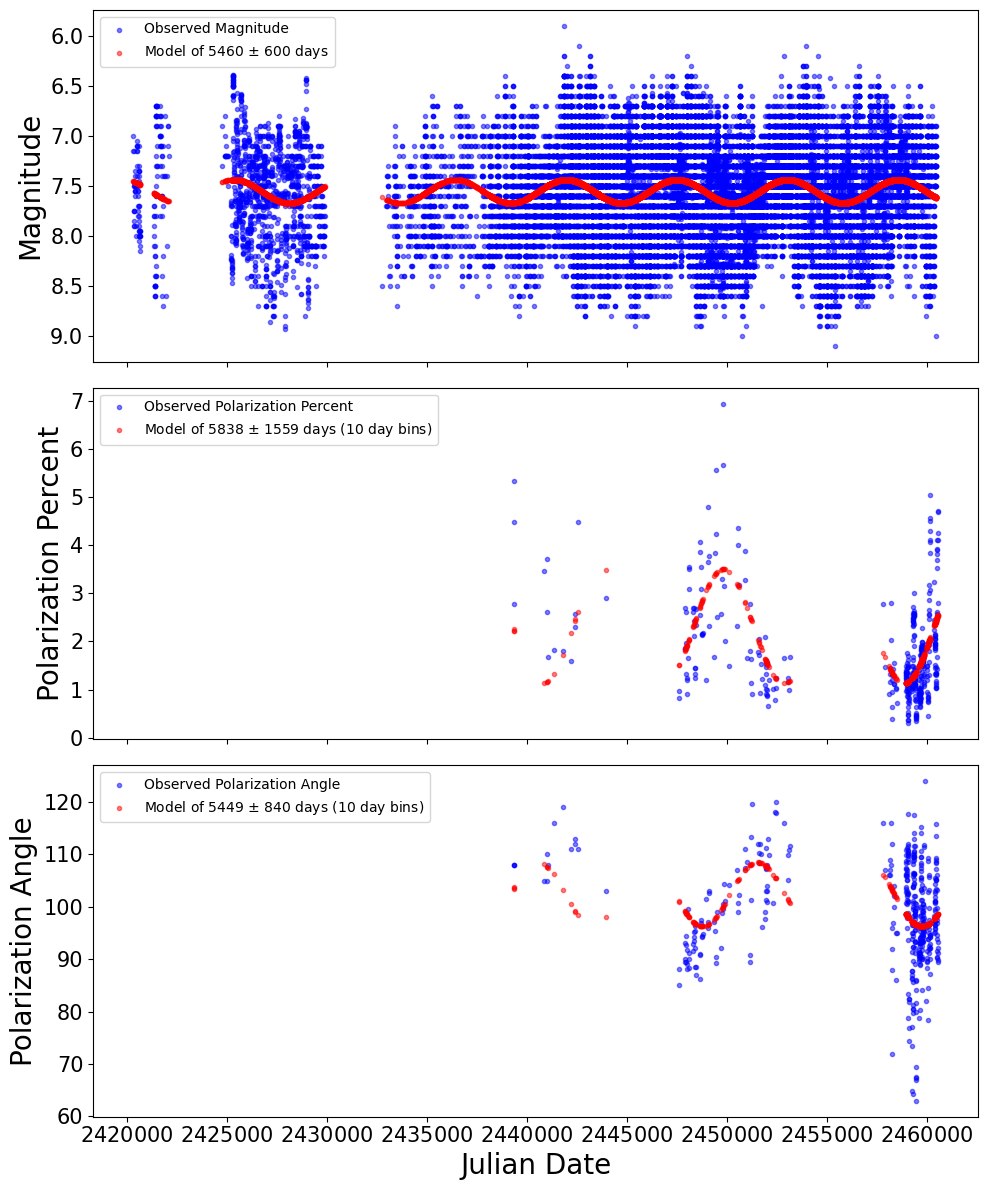}
    \caption{Result of V CVn's determined ELSP from this paper overlayed on the visual photometric data from the AAVSO (top), and the polarization percent (middle) and polarization angle (bottom) from \cite{Neilson_2014, neilson2023multiyear}.}
    \label{fig:PolPlot}
\end{figure*}

    As an additional check for purposes of planning future observation projects on V CVn, we performed a period analysis using VStar of the polarization angle and polarization percent from \cite{neilson2023multiyear}. Because the polarization data is sparse with large gaps, the results have large uncertainties, and are not conclusive. After importing the polarization data to VStar, we determine a 5838 $\pm$ 1559 day period for the polarization percent, and a 5449 $\pm$ 840 day period for the polarization angle. The photometric, polarization percent, and polarization angle data and the corresponding ELSP models are displayed in Fig. \ref{fig:PolPlot}. Though the models of polarization ELSP are not conclusive, it is worth noting that a maximum polarization percent loosely corresponds to a minimum polarization angle with a lag of around 2000 JD. As the plot indicates, V CVn might be approaching a polarization percent maximum over the next seven years, and has just undergone polarization angle minimum. Based on this trend, future observation missions can be planned to determine the credibility of this theory.

%% Please use the acknowledgment and contribution environments. This will 
%% be anonomyized when the "anonymous" style option is used. 

\section{Conclusion and Future Work}

Throughout this paper, we have analyzed light curves of AAVSO visual photometric data of V CVn, $L_2$ Pup, AK Peg, RX Boo, and Z UMa. We then constructed models of LPV pulsation for 1.0, 1.1, 1.2, 1.3, and 1.5 $M_{\odot}$ stars using MESA and GYRE to determine the pulsation periods in the AGB region and calculate the cross-harmonic periods. 

As a result of analyzing the light curves, we determine that two stars, V CVn and Z UMa, share a feature we title the Extra-Long Secondary Period, or ELSP at 5441 $\pm$ 680 days and 4668 $\pm$ 223 respectively. For V CVn, \cite{kiss2000multiperiodicity} have argued that this period is caused by cross-harmonics between the two main periods. \cite{suchko1980periodicities} argue the same case for Z UMa, which displays two closely-spaced periods as in V CVn. However, the two stars differ in the relative powers of the alleged cross-harmonic peaks. For Z UMa, the relative powers reside within 17$\%$ of each other, whereas the ELSP peak in V CVn is over 300 $\%$ stronger than the shorter cross-harmonic period. We argue that we would expect the two peak powers to be near each other if V CVn's ELSP was purely caused by cross-harmonics and not another extrinsic or intrinsic process.

While we do not find evidence for an ELSP caused by cross harmonics in our models for V CVn and Z UMa, more rigorous modeling is needed for a definite conclusion. This would include running models for higher masses, testing different pulsation mechanisms, and varying the orders of Equation \ref{eq:crossHarm}.

\begin{acknowledgments}
We acknowledge with thanks the variable star observations from the AAVSO International Database contributed by observers worldwide and used in this research.
\end{acknowledgments}

\begin{contribution}
%%This section gives authors the space to recognize author contributions. The text inside this environment is NOT counted towards the total word quanta. At a minimum, manuscripts are expected to include this text:
GMS was responsible for writing and submitting the manuscript, data visualization, and analysis.
AC was responsible for data visualization and analysis.
NS obtained part of the data used in this paper and provided feedback on the manuscript.
RI obtained part of the data used in this paper and provided feedback on the manuscript
HRN supervised the research, came up with the initial research concept, and provided feedback on the manuscript.

\end{contribution}

\software{MESA r23.05.1 \citep{paxton_modules_2011,paxton_modules_2013,paxton_modules_2015,paxton_modules_2018,paxton_modules_2019,jermyn_modules_2023, paxton_2023_7983526}
GYRE
\citep{townsend_gyre_2013}
matplotlib
\citep{Hunter:2007}
}

\appendix

\section{Linear Non-Adiabatic Prediction}
\label{app: NASolution}
The result of running a 1.3 $M_{\odot}$ GYRE pulsation modes model with a linear non-adiabatic solution; as displayed by the mode mixing, we believe the model could not accurately converge on a solution.

\begin{figure*}[h]
    \centering
    \includegraphics[scale=0.35]{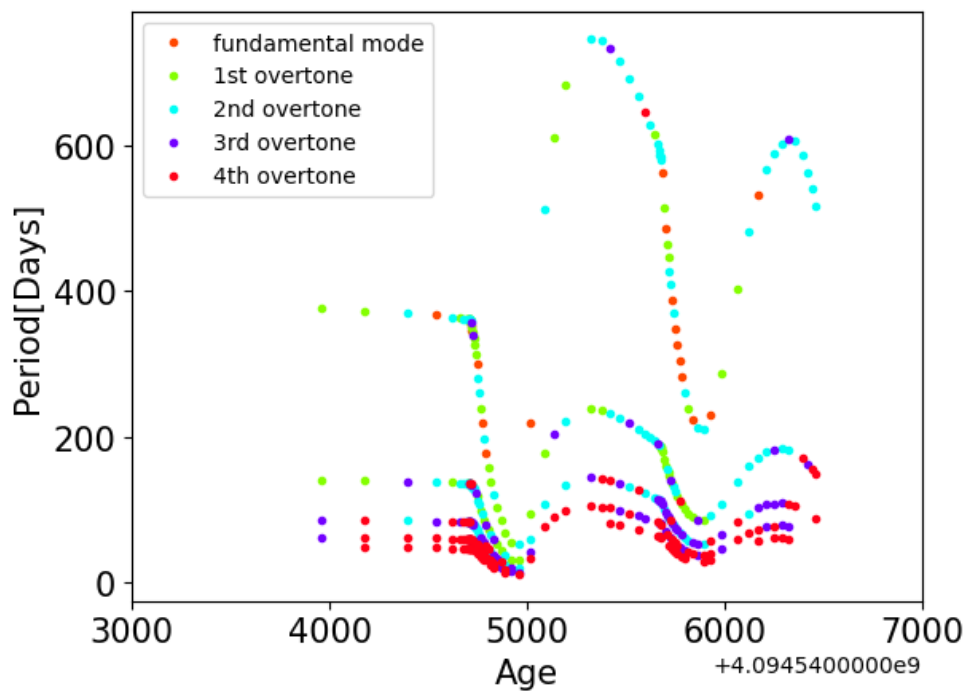}
    \caption{Linear non-adiabatic GYRE solution of a 1.3 $M_{\odot}$ star in the AGB region.}
\end{figure*}

\newpage

\bibliography{sample701}{}

\begin{thebibliography}{}
\expandafter\ifx\csname natexlab\endcsname\relax\def\natexlab#1{#1}\fi
\providecommand{\url}[1]{\href{#1}{#1}}
\providecommand{\dodoi}[1]{doi:~\href{http://doi.org/#1}{\nolinkurl{#1}}}
\providecommand{\doeprint}[1]{\href{http://ascl.net/#1}{\nolinkurl{http://ascl.net/#1}}}
\providecommand{\doarXiv}[1]{\href{https://arxiv.org/abs/#1}{\nolinkurl{https://arxiv.org/abs/#1}}}

% type= article
\bibitem[{T.~R. Bedding {et~al.}(2002)Bedding, Zijlstra, Jones, Marang, Matsuura, Retter, Whitelock, \& Yamamura}]{bedding2002light}
Bedding, T.~R., Zijlstra, A., Jones, A., {et~al.} 2002, \bibinfo{title}{The light curve of the semiregular variable L2 Puppis--I. A recent dimming event from dust,} Monthly Notices of the Royal Astronomical Society, 337, 79

% type= article
\bibitem[{D. Benn(2012)Benn}]{benn2012algorithms+}
Benn, D. 2012, \bibinfo{title}{Algorithms+ Observations= VStar,} Journal of the American Association of Variable Star Observers (JAAVSO), 40, 852

% type= article
\bibitem[{J.~R. Buchler {et~al.}(2004)Buchler, Koll{\'a}th, \& Cadmus~Jr}]{buchler2004evidence}
Buchler, J.~R., Koll{\'a}th, Z., \& Cadmus~Jr, R.~R. 2004, \bibinfo{title}{Evidence for low-dimensional chaos in semiregular variable stars,} The Astrophysical Journal, 613, 532

% type= article
\bibitem[{R.~R. Cadmus(2024)Cadmus}]{cadmus2024long}
Cadmus, R.~R. 2024, \bibinfo{title}{The Long-term Photometric Behavior of 39 Semiregular Variable Stars,} The Astronomical Journal, 167, 200

% type= book
\bibitem[{M. Catelan \& H.~A. Smith(2015)Catelan \& Smith}]{catelan_pulsating_2015}
Catelan, M., \& Smith, H.~A. 2015, Pulsating {Stars} (John Wiley \& Sons)

% type= article
\bibitem[{S. Ferraz-Mello(1981)Ferraz-Mello}]{ferraz1981estimation}
Ferraz-Mello, S. 1981, \bibinfo{title}{Estimation of periods from unequally spaced observations,} The Astronomical Journal, 86, 619

% type= article
\bibitem[{G. Foster(1995)Foster}]{foster1995cleanest}
Foster, G. 1995, \bibinfo{title}{The cleanest Fourier spectrum,} The Astronomical Journal (ISSN 0004-6256), vol. 109, no. 4, p. 1889-1902, 109, 1889

% type= article
\bibitem[{G. Foster(1996)Foster}]{foster1996wavelets}
Foster, G. 1996, \bibinfo{title}{Wavelets for period analysis of unevenly sampled time series,} Astronomical Journal v. 112, p. 1709-1729, 112, 1709

% type= book
\bibitem[{G. Foster(2010)Foster}]{foster2010analyzing}
Foster, G. 2010, Analyzing Light Curves: A Practical Guide (Lulu. com)

% type= article
\bibitem[{J.~A. Goldberg {et~al.}(2024)Goldberg, Joyce, \& Moln{\'a}r}]{goldberg2024buddy}
Goldberg, J.~A., Joyce, M., \& Moln{\'a}r, L. 2024, \bibinfo{title}{A Buddy for Betelgeuse: Binarity as the Origin of the Long Secondary Period in $$\backslash$alpha $ Orionis,} arXiv preprint arXiv:2408.09089

% type= article
\bibitem[{J.~D. Hunter(2007)Hunter}]{Hunter:2007}
Hunter, J.~D. 2007, \bibinfo{title}{Matplotlib: A 2D graphics environment,} Computing in Science \& Engineering, 9, 90, \dodoi{10.1109/MCSE.2007.55}

% type= article
\bibitem[{A.~S. Jermyn {et~al.}(2023)Jermyn, Bauer, Schwab, Farmer, Ball, Bellinger, Dotter, Joyce, Marchant, Mombarg, Wolf, Sunny~Wong, Cinquegrana, Farrell, Smolec, Thoul, Cantiello, Herwig, Toloza, Bildsten, Townsend, \& Timmes}]{jermyn_modules_2023}
Jermyn, A.~S., Bauer, E.~B., Schwab, J., {et~al.} 2023, \bibinfo{title}{Modules for Experiments in Stellar Astrophysics ({MESA}): Time-dependent Convection, Energy Conservation, Automatic Differentiation, and Infrastructure,} The Astrophysical Journal Supplement Series, 265, 15, \dodoi{10.3847/1538-4365/acae8d}

% type= article
\bibitem[{M. Joyce {et~al.}(2024)Joyce, Moln{\'a}r, Cinquegrana, Karakas, Tayar, \& Tarczay-Neh{\'e}z}]{joyce2024stellar}
Joyce, M., Moln{\'a}r, L., Cinquegrana, G., {et~al.} 2024, \bibinfo{title}{Stellar Evolution in Real Time II: R Hydrae and an Open-Source Grid of >3000 Seismic TP-AGB Models Computed with MESA,} The Astrophysical Journal, 971, 186, \dodoi{10.3847/1538-4357/ad534a}

% type= article
\bibitem[{P. Kervella {et~al.}(2016)Kervella, Homan, Richards, Decin, McDonald, Montargès, \& Ohnaka}]{kervella_alma_2016}
Kervella, P., Homan, W., Richards, A. M.~S., {et~al.} 2016, \bibinfo{title}{{ALMA} observations of the nearby {AGB} star {L} $_{\textrm{2}}$ {Puppis}: {I}. {Mass} of the central star and detection of a candidate planet,} Astronomy \& Astrophysics, 596, A92, \dodoi{10.1051/0004-6361/201629877}

% type= article
\bibitem[{L.~L. Kiss {et~al.}(2000)Kiss, Szatm{\'a}ry, Szab{\'o}, \& Mattei}]{kiss2000multiperiodicity}
Kiss, L.~L., Szatm{\'a}ry, K., Szab{\'o}, G., \& Mattei, J. 2000, \bibinfo{title}{Multiperiodicity in semiregular variables-II. Systematic amplitude variations,} Astronomy and Astrophysics Supplement Series, 145, 283

% type=
\bibitem[{B.~K. Kloppenborg(2023)Kloppenborg}]{aavso_data}
Kloppenborg, B.~K. 2023, Observations from the AAVSO International Database, \url{https://www.aavso.org}

% type= article
\bibitem[{F. Lykou {et~al.}(2015)Lykou, Klotz, Paladini, Hron, Zijlstra, Kluska, Norris, Tuthill, Ramstedt, Lagadec, Wittkowski, Maercker, \& Mayer}]{lykou_dissecting_2015}
Lykou, F., Klotz, D., Paladini, C., {et~al.} 2015, \bibinfo{title}{Dissecting the {AGB} star {L} $_{\textrm{2}}$ {Puppis}: a torus in the making \textit{({Corrigendum})},} Astronomy \& Astrophysics, 581, C2, \dodoi{10.1051/0004-6361/201322828e}

% type= article
\bibitem[{A.~M. Magalhaes {et~al.}(1986)Magalhaes, Coyne, Codina-Landaberry, \& Gneiding}]{magalhaes1986polarimetric}
Magalhaes, A.~M., Coyne, G., Codina-Landaberry, S., \& Gneiding, C. 1986, \bibinfo{title}{Polarimetric evidence for an evolving circumstellar cloud in L2 Puppis,} Astronomy and Astrophysics (ISSN 0004-6361), vol. 154, no. 1-2, Jan. 1986, p. 1-7. Research supported by the Fundacao de Amparo a Pesquisa do Estado de Sao Paulo., 154, 1

% type= article
\bibitem[{H. Neilson {et~al.}(2023)Neilson, Steenken, Simpson, Ignace, Shrestha, Erba, \& Henson}]{neilson2023multiyear}
Neilson, H., Steenken, N., Simpson, J., {et~al.} 2023, \bibinfo{title}{A multiyear photopolarimetric study of the semi-regular variable V CVn and identification of analog sources,} Astronomy \& Astrophysics, 677, A96

% type= article
\bibitem[{H.~R. Neilson {et~al.}(2014)Neilson, Ignace, Smith, Henson, \& Adams}]{Neilson_2014}
Neilson, H.~R., Ignace, R., Smith, B.~J., Henson, G., \& Adams, A.~M. 2014, \bibinfo{title}{Evidence of a Mira-like tail and bow shock about the semi-regular variable V CVn from four decades of polarization measurements,} Astronomy \&; Astrophysics, 568, A88, \dodoi{10.1051/0004-6361/201424037}

% type=
\bibitem[{B. Paxton(2023)Paxton}]{paxton_2023_7983526}
Paxton, B. 2023, Modules for Experiments in Stellar Astrophysics (MESA), r23.05.1 Zenodo, \dodoi{10.5281/zenodo.7983526}

% type= article
\bibitem[{B. Paxton {et~al.}(2011)Paxton, Bildsten, Dotter, Herwig, Lesaffre, \& Timmes}]{paxton_modules_2011}
Paxton, B., Bildsten, L., Dotter, A., {et~al.} 2011, \bibinfo{title}{Modules for Experiments in Stellar Astrophysics ({MESA}),} The Astrophysical Journal Supplement Series, 192, 3, \dodoi{10.1088/0067-0049/192/1/3}

% type= article
\bibitem[{B. Paxton {et~al.}(2013)Paxton, Cantiello, Arras, Bildsten, Brown, Dotter, Mankovich, Montgomery, Stello, Timmes, \& Townsend}]{paxton_modules_2013}
Paxton, B., Cantiello, M., Arras, P., {et~al.} 2013, \bibinfo{title}{Modules for Experiments in Stellar Astrophysics ({MESA}): Planets, Oscillations, Rotation, and Massive Stars,} The Astrophysical Journal Supplement Series, 208, 4, \dodoi{10.1088/0067-0049/208/1/4}

% type= article
\bibitem[{B. Paxton {et~al.}(2015)Paxton, Marchant, Schwab, Bauer, Bildsten, Cantiello, Dessart, Farmer, Hu, Langer, Townsend, Townsley, \& Timmes}]{paxton_modules_2015}
Paxton, B., Marchant, P., Schwab, J., {et~al.} 2015, \bibinfo{title}{Modules for Experiments in Stellar Astrophysics ({MESA}): Binaries, Pulsations, and Explosions,} The Astrophysical Journal Supplement Series, 220, 15, \dodoi{10.1088/0067-0049/220/1/15}

% type= article
\bibitem[{B. Paxton {et~al.}(2018)Paxton, Schwab, Bauer, Bildsten, Blinnikov, Duffell, Farmer, Goldberg, Marchant, Sorokina, Thoul, Townsend, \& Timmes}]{paxton_modules_2018}
Paxton, B., Schwab, J., Bauer, E.~B., {et~al.} 2018, \bibinfo{title}{Modules for Experiments in Stellar Astrophysics ({MESA}): Convective Boundaries, Element Diffusion, and Massive Star Explosions,} The Astrophysical Journal Supplement Series, 234, 34, \dodoi{10.3847/1538-4365/aaa5a8}

% type= article
\bibitem[{B. Paxton {et~al.}(2019)Paxton, Smolec, Schwab, Gautschy, Bildsten, Cantiello, Dotter, Farmer, Goldberg, Jermyn, Kanbur, Marchant, Thoul, Townsend, Wolf, Zhang, \& Timmes}]{paxton_modules_2019}
Paxton, B., Smolec, R., Schwab, J., {et~al.} 2019, \bibinfo{title}{Modules for Experiments in Stellar Astrophysics ({MESA}): Pulsating Variable Stars, Rotation, Convective Boundaries, and Energy Conservation,} The Astrophysical Journal Supplement Series, 243, 10, \dodoi{10.3847/1538-4365/ab2241}

% type= article
\bibitem[{J.~R. {Percy} \& E. {Deibert}(2016){Percy} \& {Deibert}}]{percy2016studies}
{Percy}, J.~R., \& {Deibert}, E. 2016, \bibinfo{title}{{Studies of the Long Secondary Periods in Pulsating Red Giants},} \jaavso, 44, 94, \dodoi{10.48550/arXiv.1607.06482}

% type= article
\bibitem[{N. Samus’ {et~al.}(2017)Samus’, Kazarovets, Durlevich, Kireeva, \& Pastukhova}]{samus2017general}
Samus’, N., Kazarovets, E., Durlevich, O., Kireeva, N., \& Pastukhova, E. 2017, \bibinfo{title}{General catalogue of variable stars: Version GCVS 5.1,} Astronomy Reports, 61, 80

% type= article
\bibitem[{I. Soszy{\'n}ski {et~al.}(2021)Soszy{\'n}ski, Olechowska, Ratajczak, Iwanek, Skowron, Mr{\'o}z, Pietrukowicz, Udalski, Szyma{\'n}ski, Skowron, {et~al.}}]{soszynski2021binarity}
Soszy{\'n}ski, I., Olechowska, A., Ratajczak, M., {et~al.} 2021, \bibinfo{title}{Binarity as the Origin of Long Secondary Periods in Red Giant Stars,} The Astrophysical Journal Letters, 911, L22

% type= article
\bibitem[{M. Suchko(1980)Suchko}]{suchko1980periodicities}
Suchko, M. 1980, \bibinfo{title}{The Periodicities of Z-Ursae,} Journal of the American Association of Variable Star Observers, Vol. 9, P. 74, 1980, 9, 74

% type= article
\bibitem[{R.~H.~D. Townsend \& S.~A. Teitler(2013)Townsend \& Teitler}]{townsend_gyre_2013}
Townsend, R. H.~D., \& Teitler, S.~A. 2013, \bibinfo{title}{{GYRE}: an open-source stellar oscillation code based on a new Magnus Multiple Shooting scheme,} Monthly Notices of the Royal Astronomical Society, 435, 3406, \dodoi{10.1093/mnras/stt1533}

% type= article
\bibitem[{P.~R. Wood(2000)Wood}]{Wood_2000}
Wood, P.~R. 2000, \bibinfo{title}{Variable Red Giants in the LMC: Pulsating Stars and Binaries?} Publications of the Astronomical Society of Australia, 17, 18–21, \dodoi{10.1071/AS00018}

\end{thebibliography}
\bibliographystyle{aasjournalv7}



\end{document}